\documentstyle[12pt,epsf]{article}

\def\be{\begin{equation}}
\def\ee{\end{equation}}
\def\bea{\begin{eqnarray}}
\def\eea{\end{eqnarray}}
\def\ba*{\begin{eqnarray*}}
\def\ea*{\end{eqnarray*}}
\begin{document}
\newcommand{\sheptitle}
{Equation of State of the Transplanckian Dark Energy and the Coincidence Problem }
\newcommand{\shepauthor}
{Mar Bastero-Gil$^{ (1)}$ and 
Laura Mersini$^{(2)}$}
\newcommand{\shepaddress}
{$^1$ Physics Department, Sussex University, and
\\
$^2$ Scuola Normale Superiore, Pisa, Italy}

\newcommand{\shepabstract}
{Observational evidence suggests that our universe is presently
dominated by a dark energy component and undergoing accelerated
expansion. We recently introduced a model, motivated by string theory
for short-distance physics, for explaining 
dark energy without appealing to any fine-tuning. The idea of the
transplanckian dark energy (TDE) was based on the freeze-out mechanism
of the ultralow frequency modes, $\omega(k)$ of very short distances,
by the expansion of the background universe, $\omega(k) \leq H$. 
In this paper we address the issue of the stress-energy tensor for the
nonlinear short-distance physics and explain the need to modify
Einstein equations in this regime. From the modified Einstein
equations we then derive the equation of state for the TDE model,
which has the distinctive feature of being continually
time-dependent. The explanation of the coincidence puzzle relies
entirely on the intrinsic time-evolution of the TDE equation of
state. 
} 

\begin{titlepage}
\begin{flushright}
hep-ph/0205271\\
SNS-PH/02-04\\
\today
\end{flushright}
\vspace{.1in}
\begin{center}
{\large{\bf \sheptitle}}
\bigskip \medskip \\ \shepauthor \\ \mbox{} \\ {\it \shepaddress} \\
\vspace{.5in}

\bigskip \end{center} \setcounter{page}{0}
\shepabstract



\begin{flushleft}
\hspace*{0.9cm} \begin{tabular}{l} \\ \hline {\small Emails:
mbg20@pact.cpes.susx.ac.uk,
l.mersini@sns.it} \\

\end{tabular}
\end{flushleft}

\end{titlepage}
\section{ Introduction}

Cosmological observations of large scale structure, SN1a, age of the
universe and cosmic microwave background (CMB) data, strongly indicate
that the universe is dominated by a dark energy component with
negative pressure \cite{dark}.
Besides the difficulty of coming up with a natural explanation for the
smallness of the observed dark energy, an equal challenge is the
"cosmic coincidence" problem.

Recently we proposed a model \cite{mbk} for explaining the
observed dark energy without appealing to fine-tuning or anthropic
arguments. This model is based on the nonlinear behavior of
transplanckian metric perturbation modes which was motivated by closed
string theory \cite{bfm,string1} and quantum gravity \cite{padmanabhan}. The
transplanckian dark energy (TDE) model was based on the freeze-out
mechanism of the short-distance modes with ultralow energy, by the
expansion of the background universe, $H$, and it naturally explained
the smallness of the observed dark energy.

In this paper we study the stress-energy tensor of the TDE model in
order to calculate the equation of state for these short-distance
stringy modes.  As we will show, the frozen tail modes start having a
negative pressure of the same order as their positive energy density
soon after the matter domination era. Thus it is only at low redshifts
that they become important for driving the universe into an
accelerated expansion and dominate the Hubble expansion rate $H$. A
distinctive feature of the TDE model is that its equation of state,
$w_H$, is always strongly time-dependent at any epoch in the evolution
of the universe, (e.g. $w_H =-1/3$ during radiation dominated era but
it becomes $w_H =-1/2$ at matter domination). It becomes more and more
negative at late times until it approaches the limiting value $w_H
=-1$, after the matter domination time, $t_{eq}$.

The calculation of the components of the stress-energy tensor,
$T_{\mu\nu}$, namely the pressure and energy density, is given in
Section 2. Due to the nonlinearity of short distance physics, Bianchi
identity is generically violated for all these models. Therefore one
needs to modify the Einstein equations, ($T_{\mu\nu}$), such that the
modified ones satisfy Bianchi identity.

From physical considerations, the need for modifying Einstein
equations in the nonlinear regime of short distance physics is to be
expected, due to nonequilibrum dynamics of the short distance
modes. In practical terms this is not an easy fair to carry out in an
unambiguous way, for a simple reason: we do not have a unique
effective theory valid at transplanckian energies or a lagrangian
description of the theory in this regime \cite{carlip}. The only information
available to most transplanckian models
\cite{jacobson,mbk,brand,niemeyer,akim,starobinsky,transplanck} is the field 
equation of motion (with a few exceptions like
\cite{jacobs}). Nevertheless all these models do violate Bianchi
identity and the energy conservation law, if
$T_{\mu\nu}$ is not modified accordingly.

Based on the equation of motion as our sole information for
short-distance physics, we therefore use a kinetic theory approach for
modifying Einstein equations in the absence of an effective lagrangian
description. The assumption made is that a kinetic theory description
of the cosmological fluid is valid even in the transplanckian
regime. Despite its nonlinear behavior at short distances, this
imperfect fluid shares the same symmetries, namely homogeneity and
isotropy, as the background Friedman-Robertson-Walker (FRW)
universe. Then the corrections $\tau_{\mu\nu}$ to the stress energy
tensor $T_{\mu\nu}$ will also be of a diagonal form \cite{grsean} 
\be
\tau_{\mu\nu} = (\bar{\epsilon} + \Pi) u_{\mu}u_{\nu} + \Pi g_{\mu\nu}
\ee

In Section 3 we explore the observational consequences of the model 
with the puzzle of 'cosmic coincidence' in mind. A summary is given in Section 4. A
discussion of the nonequilibrum dynamics and distribution function for
the transplanckain modes, as well as details of averaging of their
energy and pressure, are attached in the Appendix.  

\section{ The Equation of State from the Modified Einstein Equations}

\subsection{\em Analytical expression for $T_{\mu\nu}$ }

\indent Transplanckian models that investigate the sensitivity of the CMB
spectrum or Hawking radiation to short-distance physics, all introduce
a nonlinear, time-dependent frequency for the very short wavelength
modes \cite{mbk,brand,niemeyer,akim}: 
\be \omega[p] = f[p] =f[k/a] \,. \ee 
The physical momentum $p$ is related to the comoving wavenumber $k$ by $p=k/a$,
with $a$ the scale factor.  Most of these models lack a lagrangian
description and, all the information they propose about short-distance
physics is contained in the mode equation of motion\footnote{The
$\omega^2$ term collectively denotes the generalized frequency that
appears as a mass squared term in the equation. Depending in the
particular problem studied it may also include other terms to it, like
for example, the coupling of the modes to the curvature of the
universe, $a^{\prime \prime}/a$, if the equation under consideration is that of
metric perturbations.}: 
\be [\Box + \omega(k,a)^2]\phi_k = 0 \,.\ee

The expectation that Einstein equations will not hold unless they are
modified in the nonlinear regime of short distance physics is fully
reasonable and it is based on the fact that Bianchi identity and
energy conservation law will be violated due to the nonlinear time
dependence of $\omega$. In terms of kinetic theory, the time-dependence
of the group velocity $v_g$ indicates departure from equilibrium \cite{corleyjacob} 
(see Appendix). Here we study the modifications of
$T_{\mu\nu}$ for a specific class, the TDE model \cite{mbk}. Our
approach is based on kinetic theory and the pressure modifications are
obtained through balance equations.

In the TDE model we are considering, the dispersed frequency for
short-distance metric perturbation modes is: 
\bea \omega^2[p] &=& p^2 {\cal E}[p/p_c] \\ 
&=& p^2 \left[ \frac{\epsilon_1}{1+u} +
\frac{\epsilon_3 u}{(1+u)^2} \right] \label{epstein}\,, 
\\ u&=& exp[2 p/p_c] \,, 
\eea
where $p_c$ is of order the Planck mass or string scale $M$, $p$ is
the physical momentum, $\epsilon_i$ arbitrary constants.
The maximum of $\omega[p]$ is around $p \approx p_c$.  
The frequency function behaves as 
\bea \omega^2[p] &\approx& p^2 \left( 1 + O(p^2/M^2) \right) \;\;\;\, 
p \ll M \, \label{linear}
\eea
for the modes in the sub-planckian regime, and like
\bea
\omega^2[p] &\approx& \sqrt{\epsilon_1+ \epsilon_3} p^2 exp[-2p/M]
\;\;\;\, p \gg M \,. \label{exp} 
\eea 
for those modes in the TP regime. 
The nonlinear
exact function  Eq (\ref{epstein}  for the
frequency can be fitted to a good accuracy to $\omega[p]^2 \sim
\frac{p^2}{\cosh[p/p_c -1]^2}$. 

Let us refer to the wavepackets of
the modes centered around a momentum $p_i$ as particles. Then, their
group velocity $v_g = d\omega/dp$ is 
time dependent through its nonlinear p-dependence, and is different
form the phase velocity, $v_c=\omega/p$.

Lorentz invariance is broken due to the nonlinearity at short
distances. Therefore, the {\em fixed} cuttoff scale $p_c=M$, together
with all the transplanckian modes 
pick a {\em preferred frame}, the CMB frame. This frame is freely
falling along the comoving geodesics, with respect to the physical FRW
Universe\footnote{See \cite{corleyjacob} for a very nice treatment of
issues related to a fixed physical cuttoff in a preferred frame.}. 
Sometimes  we will 
refer to transplanckian modes as the modes inside a small box with
fixed Planck size, $l_p = 1/M$, in the preferred frame, since their
wavelength $\lambda_{TP} < l_p$ is smaller that the "size of the box" ,
$p>M$. In this picture, Lorentz invariance is broken in the small box
but restored in the large box with size $L=a/M$, i.e the Universe.
Thus the physical momenta modes for the "small box" bound observers in the
preferred frame are the comoving wavenumber modes for the "outside"
observers, in the Lorentz invariant FRW Universe, that "see" the
preferred frame in a free fall.

Let us address below the issue of how the energy density components 
behave with time, prior to the pressure modifications. The short
distance modes are out of thermal equilibrium, due to their nonlinear
frequency and group velocity, $v_g \ne 1$. Meanwhile a thermal state
is restored at large scales, ($\lambda \gg l_P$), where the frequency
is nearly linear and thus $v_g \simeq 1$. Thus we need to average the
contribution of the short distance modes to the energy and pressure in
the Universe, over many of their wavelengths, in order to obtain an
effective large scale thermal state.  That is why in obtaining the
equation of state, $<w_i>$ prior to the pressure modifications
$\Pi_i$, for the transplanckian modes, the averaging is done in
time-scales of cosmological order. Details of averaging are provided
in the Appendix, 5.2. The equation of state $<w_i>$ prior to viscous
pressure modifications, is obtained from the expression $<w_i> =
\frac{<\bar p_i>}{<\rho_i>}$ with $<\bar p_i> = -<\rho_i> - <(a/3)
d\rho_i /da>$ and $a$ the scale factor.   
Based on the behavior of $v_g$ with $p$, we divide the dispersion
function into 4 regions (see Fig.1)   

\begin{figure}[t]
\epsfysize=12cm \epsfxsize=12cm \hfil \epsfbox{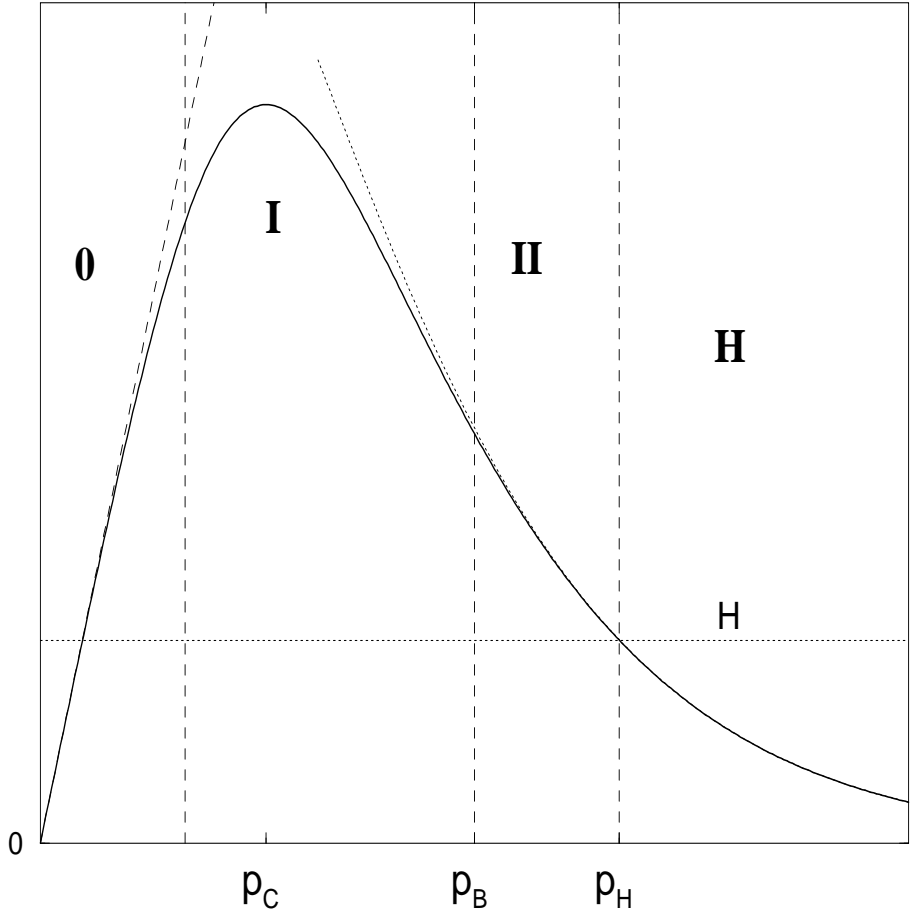} \hfil
\caption{ The dispersion function for the frequency $\omega[p]$
vs. p. The separation into 4 regions is based on the behavior of the
group velocity. The ``tail'' is denoted by region ``H''.}
\end{figure}

$\bullet$ {\bf Region 0} : Linear regime, up to $p \approx p_c$, such
that $\omega[p] \simeq p$. These modes behave as radiation, Eq.(\ref{linear}),
with the averaged pressure expression being $\bar{p}_0 = \rho_0/3$.

$\bullet$ {\bf Region I} : Around the maximum of the dispersion
function, up to some value $p_B > M$ in physical momentum, 
where $\omega$ can be expanded in a
polynomial series, and the leading order terms in Eq. (\ref{epstein})
are the first ones. Region I is dual to region 0.  
\be \omega[p]
\approx p \left( a_0 + a_1 (p/M) + a_2 (p/M)^2 + \cdot \right)\,,
\label{maximum} \ee where $a_i$ are constants, ($a_2 <0$).  We use
Eq. (\ref{maximum}) to estimate the energy density 
\bea \rho_I
&\simeq& \frac{C}{a^4} \int_M^{k_B} dk k^3 \left( a_0 + a_1 \frac{k}{a
M} + a_2 \left(\frac{k}{aM} \right)^2 + \cdots \right) \\
%
%
       & \simeq& \frac{C M^4}{a^4} \left(\frac{a_0}{4} (x_B^4- 1) +
\frac{a_1}{5a} (x_B^5 - 1) + \cdots \right) \\
      & \propto & \frac{M^4}{a^4} \,, \eea 
where $x_B= k_B/M > 1$, $\langle x_B\rangle = O(1)$. The constant $C =|\beta_p|^2$
denotes the Bogolubov coefficient squared, which in our model does not
depend on the momentum $p$ \cite{mbk}.  
Therefore, $\rho_I$ behaves like radiation plus $O(1/a^2)$
corrections in its averaged equation of state, $\langle \bar{p}_I\rangle = (1/3 -
A/a^2 +..)\langle \rho_I\rangle$ . Regions $I$ and $0$ contribute to the radiation
energy component in the Universe.

$\bullet$ {\bf Region II}: From some mode, $p_B \gg M$, onwards
defined such that its frequencies can be best fitted to an exponential
dependence on $p$, $\omega[p] \approx p exp[-p/M]$. The energy for
this region is 
\bea \rho_{II} &\simeq& \frac{C}{a^4} \int_{k_B}^{k_H}
dk k^3 exp[-2 k/aM] \\ &=& C \frac{M^4}{a^3} \left( F[x_B] - F[x_H]
\right) \,, 
\eea 
where, 
\be F[x_i] = \left( \frac{x_i^3}{2} + \frac{3
x_i^2}{4} + \frac{ 3 x_i}{4} + \frac{3}{8} \right) exp[- 2x_i / a] \,.
\ee
and $x_i = k_i / M $. Since $x_B > 1$ then $F[x_B] \simeq \frac{1}{2}
x_B^3 exp[-2 x_B / a]$. Thus $\rho_{II}$ behaves as matter when averaged
over many oscillations and its averaged pressure\footnote{See
Appendix for details of averaging.} is $\langle \bar{p}_{II}\rangle =
(-B/a)\langle \rho_{II}\rangle \simeq 0$. Also since $x_H > x_B > 1$ then $F[x_B] >> F[x_H]$.

$\bullet$ {\bf Region ``H''} : This is our ``tail''\cite{mbk}, defined
as the part of the graph for which the frequency of the modes is
smaller than the Hubble parameter, $H$. 
The functional behavior of the frequency with $p$ is the same as in
region II, therefore the averaged pressure expression for this region
is the same as that of Region $II$, that is $\langle \bar p_{H}\rangle \simeq
0$. But the lower limit of integration $k_H$ (or $p_H$) is given by
the physical condition of the freeze-out of the modes by the expansion
of the background universe, 
\be \omega_H[p_H]=H \,.  \label{kh} \ee
This region includes the modes from $p_H$ to $\infty$ in the range
where $\omega$ is exponentially suppressed. The energy density of the
tail is
\be \rho_{H} \simeq \frac{C}{a^3} \frac{k^3_H}{2M^3} exp[-2 k_H/aM]
\ee 
Notice that due to the freeze-out, the evolution of the $k_H$ mode
is highly nontrivial and thus corrections to the averaged pressure
term $\langle \bar{p}_H\rangle \simeq 0$ will be important.

Modes in the tail, between $p_H$ to $\infty$, behave
differently from the other modes, since their time-dependence is
controlled by the Hubble expansion, Eq. (\ref{kh}). On the other hand,
all modes with momenta $p \le p_H$ redshift in the same way with the
scale factor, towards decreasing values, i.e the linear
regime\footnote{Modes in the linear regime are referred to as "normal
modes". }.  Nevertheless, these regions, ($0,I,II$), also receive small
modifications to their pressure term from the deep transplanckian
regime. We show below that the modifications due to the
$p_H$-defrosting effect, are non negligible and important only in the
highly nonlinear regime, around $p_H$.

Now, we would like to estimate the corrections to pressure, $\Pi_i$,
for all these regions, with the notation, $P_i$ for the effective
modified pressure 
\be P_i \rightarrow \langle  \bar p_i \rangle + \Pi_i \,,
 \ee
where the index runs to $i =0, I, II, H$.
The averaged unmodified pressure expressions, $\langle  \bar p_i \rangle$,  are 
\bea 
\langle \bar p_{0,I}\rangle \simeq (\frac{1}{3} - A/a^2 + \cdots)
\langle \rho_{0,I}\rangle\,, \\ \langle \bar p_{II,H}\rangle \simeq (
- B/a + \cdots) \langle \rho_{II,H}\rangle 
\,. 
\eea 
The terms that go as inverse powers of $a$ can be neglected and
A,B are numerical constants  related to the averaging (see Appendix
for details). We
refer to these expressions as "bare pressures" in order to distinguish
them from the viscous pressure modifications terms (defined below).

In a similar manner to particle creation cases \cite{partcreat}  in imperfect
fluids \cite{zimdahl,ehlers}, the highly nontrivial time-dependence of the
mode $p_H$ and the transfer of energy between regions, due to the
defrosting of this mode across the boundary $p_H$, gives rise to
pressure corrections in the fluid energy conservation law. The
defrosting of the modes results in a time-dependent "particle number"
for regions near $p_H$. From kinetic theory we know that this
"particle creation", (the defrosting of the modes), gives rise to {\em
effective viscous pressure modifications} \cite{zimdahl,ehlers}. The
term $\Pi_i$ 
denotes the effective viscous pressure modification to the "bare"
pressure, $\langle \bar p_i\rangle$. 

The criteria we will use for modifying $T_{\mu\nu}$ is that Bianchi
identity must be satisfied \cite{lorentzinvar} with the new expressions for
pressure\footnote{From here on we drop the $\langle \cdots\rangle$ 
notation and denote the averaged "bare" pressure by simply $\bar p_i$
instead of $\langle \bar p_i\rangle$.}, $P_i$,
\be \Sigma_i [ \dot \rho_i + 3 H ( \rho_i + \bar{p}_i + \Pi_i )] =
\Sigma_i [ \dot \rho_i + 3 H ( \rho_i + P_i)] = 0 \,,  
\ee
with $i=0,I,II,H$. Let us write this expression explicitly in terms of
its energy and bare pressure components, and collect the contributions
of regions 0,I into one combined radiation energy, $\rho_R = \rho_0 +
\rho_I$: 
\bea 
\dot \rho_{II} + 3 H (\rho_{II} + \bar{p}_{II}) + \dot
\rho_{R} + 3 H (\rho_{R} + \bar{p}_{R}) &=& -3 H \Pi_{II} \,,
\label{rhoII} \\
\dot \rho_{H} + 3 H (\rho_{H} + \bar{p}_{H}) &=& -3 H \Pi_{H} 
\,, \label{rhoH} 
\eea 
where $\bar{p}_{II, H} \simeq 0$, $\bar{p}_R \simeq 1/3 $. 
So, we have imperfect fluids in regions II and ``H'' , and
eventually their energy is transferred, due to the redshifting effect,
to regions 0 and I, which is why these regions also receive pressure
modifications. Nevertheless, the viscous pressure corrections to the
"radiation" modes are very small since the energy and the volume of
phase space occupied by them is very large ($a^3$ times larger the
Planck size volume). These regions are in a nearly equilibrium
situations, (see  Appendix). All pressure corrections are estimated
below.

Let us find out $\Pi_i$, in order to solve Eqs. (\ref{rhoII}) and
 (\ref{rhoH}). As explained, the presence of $\Pi_i$ is due to the
 exchange of energy between the two regions, from the defrosting of
 the modes $p_H$ at the boundary. This is directly related to the time
 dependence of the boundary $p_H$, which in turn is going to be
 controlled by the Hubble parameter $H$. In essence, there is an
 exchange of modes between region $(R + II)$ and ``H''. Although the
 specific number of particles\footnote{We are loosely using the term
 particle here to refer to the wavepackets of the transplanckian
 modes, centered around a momenta $p_i$} in each of these regions,
 $N_{II}$ and $N_H$, is not conserved, their rate of change, in the physical FRW Universe, is
 related through the conservation of the total number of particles which contains both of these components 
\be
\dot N_T = 0\,, \label{dotNT}
\ee
Each component satisfies\footnote{Vector objects related to the flow direction of the fluid are denoted in bold letters, e.g. ${\bf
N_i }= N_i u_a$ with $u_a$ being the unit 4-velocity of the
fluid and the corresponding modulus of this vector being $N_i$. Notice that the factor $(3H N_H)$ in Eq. (\ref{dotNH}) is related to the fact that the preferred frame for the tail modes is falling along comoving geodesics in the Universe.}

\bea
{\bf \dot{N}_{II}}&=& \Gamma_{II} {\bf N_{II} } \,,\label{dotNII}\\
{\bf \dot{N}_{H}}&=&(3H - \Gamma_{H}) {\bf N_{H} }\,,\label{dotNH}
 \eea 
where  the "decay rates" of the regions $\Gamma_i$  account for the
 rate of change in the number of their ``particles" (modes), due to
 the defrosting effect.

The system is not yet in equilibrium. The change in the number of
"particles" is giving rise to the effective viscous pressure,
$\Pi_i$. Even prior to the freeze-out effects, that is, even for
$\Gamma_{H,II} = 0$, the short distance modes in region II and III
were out of thermal equilibrium, due to their nonlinear frequency and
group velocity, $v_g \ne 1$.

The contribution terms to pressure, $\Pi_i$, are related to $\Gamma_i$
through \cite{zimdahl,ehlers} 
 \bea 
3 H \Pi_H &=& - ( \rho_H + \bar{p}_H)
\Gamma_H \,, \label{PiH}\\ 3 H \Pi_{II} &=& - [( \rho_{II} +
\bar{p}_{II}) +( \rho_R + \bar{p}_R) ] \Gamma_{II} \label{PiII}\,.
\eea 
Therefore, Eq. (\ref{rhoH}) reduces to 
\be \dot \rho_H + (3 H -
\Gamma_H) (\rho_H + \bar{p}_H) =0 \label{dotrhoH}\,, 
\ee which can be
also recast as: 
\be \dot \rho_H = \frac{ \dot n_H}{n_H} ( \rho_H +
\bar{p}_H) \,,  
\ee
where $n_H = N_{H}/ a^3$ is the ``particle'' number
density for the region of modes from $p_H$ to infinity. The flow of
particles is described by ${\bf n_{H} }= n_H u_a$, with $u_a$ the unit 4-velocity
vector of the fluid. Notice that since the group velocity  in the H-region is is
negative, particles in this region flow in a direction $v_g$ which
opposite to the direction of their momenta, $k$.  The rate $\Gamma_H$,
calculated below, is positive, and 
the increase in the number of particles $N_H$ as given by $\Gamma_H$
does not allow the energy density of the "tail" to redshift as fast as
matter. This indicates that although small, $\rho_H$ eventually will
come to dominate the total energy density.

The number of "particles" $N_H$ contained in the tail regime, in its preferred frame, is given
by 
\bea N_H &\simeq& C_H \int_{k_H}^{\infty} dk k^2 exp[-k/aM]
\nonumber \\
& \simeq & C_H (a M) k_H^2 exp[-k_H/aM]  \label{NH} \,.  
\eea where
$C_H$ is the constant proportional to the Bogolubov
coefficient\footnote{In our model $|\beta_p|^2$ calculated in \cite{mbk}
resulted in a scale invariant spectrum. It does not depend on the wavenumber
$p$, the reason why it can be pulled out of the integral and factored
into the coefficient $C_H = N \beta_p^2$, with N an overall
normalization constant we are keeping for the sake of generality.}
$\beta_p$. We
can now calculate the energy transfer, due to the defrosting of the
modes $k_H$, between the tail region and region $II$ from the balance
equation for $N_H$, Eq. (\ref{dotNH}), where $\dot N_H$ is: 
\bea 
\dot N_H &\simeq& C_H \int_{k_H}^{\infty} dk k^2 \left( \frac{k}{aM}
\right)exp[-k/aM] - C_H k_H^2 exp[-k_H/aM] \dot k_H \nonumber \\
&\simeq& 3 H N_H - C_H k_H^2 exp[-k_H/aM]( \dot k_H - H k_H) \nonumber
\\ & \simeq & N_H \left( 3H + \frac{k_H/aM}{ k_H/aM -1} \left
( \frac{\dot H}{H} \right) \right) \,. 
\eea 
In the last line we have
used the approximation in Eq. (\ref{NH}), and: 
\be 
\frac{\dot p_H}{p_H} =  \frac{ p_H/M}{ 1 - p_H/M}( \frac{\dot H}{H} ) \label{dotpH}\,, 
\ee derived from
Eq. (\ref{kh}). When $p_H \gg M $ (which always holds), we have
for $\Gamma_H$: 
\be \Gamma_H \simeq 3H + \frac{\dot H}{H} \simeq 3 H
(\frac{1 - w_{total}}{2})\,, \label{gammah} 
\ee where 
$w_{total}=\bar{p}_{total}/\rho_{total}$. Therefore, when $w_{total} \rightarrow
-1$ then $\Gamma_H$ reaches its limit, $\Gamma_H \rightarrow 3H$.
$\Gamma_H$ can not change anymore once this limit is reached because
the Hubble constant and the mode $p_H$ freeze to a time-independent
value. Notice that $\Gamma_H$ is positive for
all equations of state $w_{total} \le 1$ and thus it slows down the
dilution of the tail with the scale factor.

We can repeat the same procedure for the modes in region $0,I,II$ in
order to obtain a closed equation for $\dot \rho_{R,II}$ , similar to
Eq. (\ref{dotrhoH}), i.e. that is given entirely in terms of
$\rho_{R,II}$ and $\bar{p}_{R,II}$
\be \dot \rho_{II,R} + (3 H
-\Gamma_{II}) (\rho_{II,R} + \bar{p}_{II,R}) =0 \,, 
\ee 
where we have used Eq. (\ref{PiII}).

Let us now try to relate $\Gamma_H$ to $\Gamma_{II}$.  
The total number of not frozen particles, $N_{II}$, in the region from
zero to $k_H$ is given by
\be 
N_{II}  \simeq  C_H [4 M^3 +  M p_H \omega_H ] \label{N2} \,.  
\ee

From the total balance equation for the particle number between the
two regions, $(R + II)$ and region ``H'' in the comoving volume, we have $
{\bf \dot N_{total}} = 0$, where  ${\bf \dot N_{II}} = \Gamma_{II}
{\bf N_{II}}$ and ${\bf N_H} = M p_H^2 Exp(-p_H /M) u = M p_H H u$. Thus 

\be   \Gamma_{II}= \frac{M p_H H}{4 M^3 + M p_H H}(\Gamma_{H} - 3H)  \,, 
\ee
and $N_T = N_{II} - C_H (M p_H \omega_H) = 4 C_H M^3$. In obtaining
the scalar quantity for the number of particles $N_T$ from their flow
${\bf N_T}$ , the negative sign picked up in the second term in $N_T$
is related to the fact that the flow of the tail's defrosted modes is
in the opposite direction to their momenta, due to their negative
group velocity. 
Therefore, by plugging in the expression of $N_{II}$ from Eq.
(\ref{N2}), we get that in the limit $N_{II} \gg M p_H H$,
$\Gamma_{II}$ is smaller than $\Gamma_H$  and negative, given by
the expression: 
\be \Gamma_{II} = - (3H - \Gamma_H) \frac{p_H}{4 M}
\frac{H}{M} \,. \label{gamma2}
\ee
Since $p_H H^2 << M^2 H$ then $y=
\frac{p_H H}{M^2} \simeq O(H/M)$ is going to be much less than 1 for
as long as the expansion is not dominated by the tail. From the
condition $\omega_H[p_H]=H$, and the time evolution of the
physical momentum $p_H$ in Eq. (\ref{dotpH}),
we have that $\dot p_H / p_H
\rightarrow - \dot H / H$, when $p_H \gg M$.  The exact value of $y$
does not matter and it is small. The tail
domination case, when $\rho_H$ becomes comparable to $\rho_{II}$,
should be treated separately since 
$\Gamma_{II} \rightarrow 0$. The pressure modification $\Pi_{II}$
increases the dilution of $\rho_{II}$ as determined by the equation
for $\dot \rho_{II}$. This equation shows that due to the modified pressure effects, $\rho_{II}$ goes to
zero faster than a matter energy density component.

\subsection{Equation of State}

In this part we calculate the effective equation of state for all the regions,
from the pressure expressions, $\bar{p}_i$, $\Pi_i$, that were
obtained in the previous section. Starting with region ``H'', we have
\bea 
\frac{\dot \rho_H}{\rho_H} &=& (\Gamma_H - 3H) ( 1 + w_H) = -\frac{ 3 H}{2} (1 + w_{total}) (1
+ w_H) \, 
\eea
where we have defined
\be
\frac{\dot H} {H} = -\frac{3}{2} ( 1 + w_total) \,,
\ee
with $w_{total}$ referring to the effective equation of state for the
total energy density. 
The ``effective" equation of state for the tail can be read from this
expression to be 
\be 1+ \tilde{w}_H= \frac{1}{2} ( 1 + w_{total}) (1
+ w_H) \,,  \label{wH}
\ee
with $w_H \simeq 0$. The time evolution for $\rho_H$ is 
\bea 
\rho_H &=& \rho_H(0) exp[ - \frac{3}{2} \int (1 + w_{total})(1+w_H) d \ln a]
\\ &=& \rho_H(0) exp[ - 3\int (1+\tilde{w}_H) d \ln a ] \,. 
\eea
During radiation domination, $w_{total}=1/3$, then the effective
equation of state for the tail is $\tilde{w}_H= -1/3$;  for matter 
domination, $w_{total}=0$, then $\tilde{w}_H=-1/2$; at the start of
the accelerated expansion, $q=0$, $\rho_{II} = \rho_H$, we have
$w_{total} = -1/3$, $\tilde w_H = -2/3$; and finally, if
the tail dominates, then the only solution to the Friedman equation is
given by $w_{total}\simeq \tilde{w}_H$, with
$\tilde{w}_H=-1$. Therefore, the tail starts behaving as dark energy
only recently, when its equation of state $\tilde w_H$ becomes close
to the limiting value, $\tilde w_H = -1$.

By the same procedure, we can now estimate the effective equation of
state for $\rho_{II}$ in terms of $\Gamma_{II}$: 
\bea 
%
\frac{\dot
\rho_{II}}{\rho_{II}} &=& \Gamma_{II} - 3H = -3H [ 1 +
(\frac{1-w_{total}}{2}) y ] \, \label{rhodot2} 
\eea Thus the effective equation of
state for region II, obtained from Eq. (\ref{rhodot2}) is: 
\be 1+ \tilde{w}_{II}=
( 1 + y \frac{1-w_{total}}{2 }) (1 + w_{II}) \,, \label{wII}
\ee 
where $y =
(\frac{p_H}{M}) (\frac{H}{M}) = O(H/M)$ and $w_{II} \simeq 0$.

The time evolution of the $\rho_{II}$ energy density component is \be
\rho_{II} = \rho_{II}(0) exp[ - 3\int (1+\tilde{w}_{II}) d \ln a ] \,,
\ee
In a radiation dominated universe, $H=M/a^2$ therefore $(H/M) =
(1/a^2)$, and matter dominated, $H = M/a^{3/2}$, so then $(H/M)
=(1/a^{3/2})$. The point is that the correction $y$ to the matter
equation of state for region $II$ is really small for up to the
equality time. During most of the history of the Universe, $y$ goes as
an inverse power of the scale factor a. So region II behaves pretty
much like matter. The special era when the tail eventually
dominates the expansion and H becomes a constant, (at $p_H=constant$,
$\tilde w_H=w_{total}=-1$) is discussed below in Sect.3.

Similarly, by repeating the same steps, from Eq. (\ref{rhoII}), it can be shown
that the modifications to pressure for $\rho_R$, regions 0,I, are very
small indeed. Thus their effective equation of state remains very
nearly that of radiation, $\tilde w_{R} =1/3$. In order to avoid
repetition, we will not carry out the calculation for the effective
equation of state of $\rho_R$, as the procedure is essentially the
same as for $\rho_{II}$, and it results in an effective radiation
equation of state 
\be 1+ \tilde{w}_{R}= ( 1 + y \frac{1-w_{total}}{2
}) (1 + w_{R}) \,, 
\ee

\section{ The Issue of Coincidence and Comparison to Observation}

From the computation of the pressures $\bar{p}_H$ and $\Pi_i$, it is
clear that the initial radiation is redshifted faster than the other
components of the total energy density. We can ask the question at
what time, $t_{eq}$, the $\rho_{II}$ components of matter becomes
comparable to radiation 
\be \rho_R \simeq \rho_{II} \simeq
\frac{\rho_{total}}{2} = \frac{3}{2} H_{eq}^2 M^2 \,, 
\ee with 
\be
\rho_{II} \simeq C p_B^3 exp[-2 p_B/M] \simeq \frac{3}{2} H_{eq}^2
M^2 \,.
\ee 
From the above equation we obtain $p_B$ at $t_{eq}$: 
\be p_B \simeq \frac{M}{2} \ln \left[\frac{2 C p_B^3}{3 H_{eq}^2 M^2}
\right]\,. \label{kb} 
\ee 
On the other hand, from Eq. (\ref{kh}) we
have: 
\be p_H \simeq M \ln \left[ \frac{p_H}{H} \right] \label{kh2}
\,, 
\ee 
and comparing Eqs. (\ref{kb}) and (\ref{kh2}), it is
clear than $p_B(t_{eq}) < p_H(t_{eq})$, and therefore 
\be \rho_H (t_{eq}) < \rho_{II}(t_{eq}) \,.  
\ee 
So matter-radiation equality
takes place well before the eventual ``tail'' domination.  From the
equations of state, $\tilde{w}_H$ and $\tilde w_{II}$, Eqs. (\ref{wH})
and (\ref{wII}), we have that $\rho_{II}$ always dilutes faster than $\rho_H$. Thus
the inequality in Eq. (60) holds true not only at $t=t_{eq}$ but at
all earlier times before $t_{eq}$. Generally, there may be other
sources of matter and radiation in the Universe, besides the
contribution from the transplanckian modes. Although these components
would not be affected by the viscous pressure corrections, $\Pi_i$,
their contribution should be including in the Friedman equation when
determining the equality time, $a(t_{eq}) =a_{eq}$. Since their effect
to the expansion is well studied and known, here we chose to
focus our attention only on the role of the transplanckian modes.

Let us now estimate the time at which the tail takes over to dominate
the expansion and address the issue of the cosmic coincidence.
As we will see below, the effective equation of state $\tilde{w}_{II}$
for $\rho_{II}$ changes from $\tilde{w}_{II}\simeq 0$ to
$\tilde{w}_{II} \ge 0$. Let us start by
asking at what time, $a_{DE}$, we have 
\be \rho_H = \rho_{II} =
\rho_{total}/2 
\ee
or in terms of the density parameters $\Omega_H=\Omega_{II}$. From the Friedmann Eq. for the expansion
and the relation of $\tilde{w}_{H}$ to $w_{total}$ it is
straightforward to find out that at $a=a_{DE}$ we have: 
\be
\tilde{w}_H (a_{DE}) = -\frac{2}{3} \,,\;\;\;\;w_{total}(a_{DE}) =
-\frac{1}{3}\,, \ee and therefore 
\be a_{DE}= \left(
\frac{\rho_{II}^{(0)}}{\rho_{H}^{(0)}}\right)^{2/3(w_{total}-1)} =
\left( \frac{\rho^{(0)}_H}{\rho^{(0)}_{II}} \right)^{1/2}= \left (
\frac{\Omega^{(0)}_H}{a_{eq}^3}\right)^{1/2} \,.  
\ee
with $\rho^{(0)}_i$ being the value of the i-th component at equality
time, $a_{eq}$. It is interesting to notice that $w_{total}=-1/3$
corresponds to the transition time where the deceleration parameter,
\be q \simeq \frac{1}{2}( 3 w_{total}+1)\,, \ee changes sign and goes
through zero.This means that acceleration starts at the same time
$a_q$ as the dominance of the tail, $a_{DE}$, i. e.,
$a_q=a_{DE}$. Using the Friedman expansion law, we can find the
solution for the scale factor $a$, after the time $a_q$: 
\be \left
( \frac{\dot a}{a}\right)^2 = \frac{\rho^{(0)}_{II}}{3 M^2} +
\frac{\rho^{(0)}_{H}}{3 M^2} = \frac{H^2(a_q)}{2} \left
[ \left(\frac{a_q}{a}\right)^2 + \frac{a_q}{ a} \right]\,.  \ee
Therefore, 
\be \int_1^{a/a_q} \sqrt{\frac{a/a_q}{(a/a_q)^2+1}}
d(a/a_q) = \frac{H(a_q)}{\sqrt{2}} (t -t_q) \,.  
\ee
This integral can be done exactly and it is messy. The important point
about it is that it gives a power law accelerated expansion,
$\frac{a}{a_q} \simeq t^n$ with $n\ge 2$.
Clearly the tail is behaving as dark energy and it is dominating the
expansion soon after $a_q$.

Let us understand physically what is happening around the time $a=a_q$,
and why $a_q=a_{DE}$.  As showed in Sect. 2, due to the strong coupling
of the tail evolution to the Hubble constant, $H$, and therefore a
coupling to $\rho_{total}$, $\tilde w_H$ becomes more and more
negative with decreasing values of $w_{total}$, soon after
$a_{eq}$. Thus the tail starts behaving as dark energy, dominates the
expansion, and approaches its limiting value $\tilde w_H =-1$ only at
late times, when other energy contributions to $\rho_{total}$ become
negligible.  From the Eqs. (\ref{gammah}) and (\ref{gamma2}), when
$\rho_{II} \simeq \rho_H$ we obtain $\Gamma_{II} \simeq
- 2 y \Gamma_H$. When the tail comes to
dominate, the only solution to the Friedman equation is $\Gamma_{H}
\rightarrow 3 H$, and then $w_{total}= \tilde{w}_H \simeq -1$. This
means that the tail dominates the expansion ($w_{total}=-1$) very fast
and, around that time $\rho_{II}$ has become nearly zero due to the
$\Gamma_{II}$ viscous pressure corrections. Recall, that in this
estimation we assumed that 
around the time $a_q$, $\rho_{II}$ is the only source of matter. The
fast dilution of $\rho_{II}$ as compared to $\rho_H$, due to the viscous pressure effects of
$\Gamma_{II}$,
is the reason why $q=0$ occurs at the same time as the tail dominance,
$a_{DE}$. With no other sources of matter, $w_{total}$ almost
immediately goes from $w_{total}=-1/3$ to $w_{total}=-1$. Thus
$a_{DE}$ has occurred very recently indeed.  If we consider other
matter contributions in the Friedman equation, that are independent of
$\rho_{II}$, $\Gamma_i$, then the time the expansion takes between the
start of the accelerated expansion $a_q$, and the time of tail dominance time over 
$\rho_{total}$, (i.e. when $w_{total}= w_H \simeq -1$) becomes a bit
longer. This time interval from $a_q$ to present is also the time
interval when the tail has acquired a dark energy equation of state,
until it reaches its limiting value, $\tilde w_H = -1$.

Therefore, {\em cosmic coincidence is explained naturally by the intrinsic
time evolution} of the effective equation of state for the tail,
{\em $\tilde w_H$ }.

\section{ Summary}

Models of nonlinear short distance physics discussed recently in
literature \cite{brand,niemeyer,akim,starobinsky,transplanck,jacobs,corleyjacob},
usually introduce a time dependent frequency, at 
the level of the equation of motion for the field. As a result Bianchi
identity is generically violated which indicates that Einstein
equations need to be modified in the high energy regime. It is
difficult to do so without an effective lagrangian description of the
theory in the transplanckian (TP) regime.

We therefore used a kinetic theory approach, in order to estimate the
short distance modification to the cosmic fluid stress energy tensor
for the model of \cite{mbk}. It is not clear to us whether this
procedure determines the modifications in a unique unambiguous way, or
whether fluid idealization and the assumption that kinetic theory
remains valid at such high wavenumber modes, is a good
approximation. Nevertheless, we believe that without an effective
lagrangian, kinetic theory is the only available tool to get some
sensible results for the contribution that TP modes have in the long
wavelength regime.

In a previous paper \cite{mbk}, we showed that the energy contribution
of the tail modes is comparable in magnitude to the observed dark
energy in the universe. In this work we calculated the effective
equation of state, $\tilde w_H$, for these tail modes in order to
address the cosmic coincidence issue and showed that the tail modes
behave as dark energy only at late times.

The tail has an exponentially suppressed frequency and all the modes
with $\omega<H$ are frozen out by the expansion of the background
universe. However, due to the short distance pressure modification,
the tail does not always behave as dark energy. The highly nontrivial
time-dependence of tail's dominant mode $p_H$, tracks the evolution of
the total energy density $\rho_{total}$ through its strong dependence
on the Hubble constant, $H$. The dependence of $p_H$ on $H$ is given
by the freeze-out condition.

As a result of the coupling of the $p_H$ mode to $\rho_{total}$, the
 tail equation of state $\tilde w_H$ follows the evolution of
 $w_{total}$, such that $\tilde w_H \simeq \frac{(w_{total}
 -1)}{2}$. For this reason, $\tilde w_H$ acquires more and more
 negative 
 values as $w_{total}$ decreases, from radiation to matter. The tail
 has a slower dilution with the scale factor compared to the other
 components and it starts dominating the expansion and behaving as
 dark energy only recently, from the time when the deceleration
 parameter $q$ changes sign at time $a_q$.  From this point, $a_q$, (
 with $w_{total} =-1/3$, $\tilde w_H = -2/3$) and onwards, the tail
 drives the Universe into an accelerated expansion, and soon reaches
 its limiting value of $ w_{total} \simeq \tilde w_H \simeq -1$ with
 $\rho_{total} \simeq \rho_H $. Therefore, cosmic coincidence in this
 model is explained naturally from the time evolution of the tail's
 $\tilde w_H = f[w_{total}]$. This is the most important result of
 this paper.

The TDE model was motivated by closed string theory of Brandenberger-Vafa model \cite{string1}. Therefore its observational implications may be explored as indirect string signatures.
Some of the distinctive features of the TDE model are the predictions
that\footnote{These predictions are in the absence of other matter
sources.}: the change in sign of the deceleration parameter, $q=0$,
occurs at the same time as the start of tail dominance, ie. the time
when the tail energy is half of the total, $a_q =a_{DE}$; this
accelerated expansion occurs very recently; due to the viscous
pressure effects, the matter contribution from the TP modes goes
through a change of its equation of state such that it starts
decaying faster than normal matter. The latter effect shortens the time the Universe takes to
change the parameters from $w_{total} \simeq -1/3, q=0$ to the time
when $w_{total} \simeq -1,\rho_{total} \simeq \rho_H $. To get these
numbers, we ignored other matter sources and also the short distance
corrections to the equations of state $\langle w_i\rangle$ that go as inverse
powers of the scale factor, $O(1/a^n)$. Perhaps these corrections may
be important recently, in delaying the time it takes the tail to
behave as dark energy, $\tilde w_H \simeq -1$. 
These features can be  scrutinized with respect to observation \cite{eos}. 

{\em Note Added} : 
Results obtained in this paper for the tail's equation of state,
$\tilde w_H$, differ from those reported in Lemoine et al. \cite{lemoine}.

The fact that we include the curvature term $a^{\prime \prime} / a$ under the
definition of the generalized frequency $\omega^2$, while they keep the
two contributions separate, is not the only source of discrepancy.
There are fundamental differences between the two studies. Authors of
 \cite{lemoine} ignore the crucial effects of the out-of-equilibrium
dynamics and the breaking of Lorentz invariance, by the nonlinear short
distance modes, when carrying out their calculation for $\rho_H$ and $p_H$. In
their approach these effects would be contained in the dynamics of the
vector field, $u_\mu$ described by a lagrangian ${\cal L}_u$ for this field.
Obviously, the choice of $u_{\mu}$ field dynamics and its Lagrangian ${\cal
L}_u$ {\em
strongly depend on the details of the short distance nonlinear model
considered}.
The expression for ${\cal L}_u$ these authors borrow from the
Corley-Jacobson (CJ) model \cite{jacobson} is {\em consistent only with
the dispersion function of the CJ model} for stationary backgrounds, since
both ${\cal L}_{cor}$ and the terms given by ${\cal L}_u$ contain up to 4-th
order derivatives, together with the antisymmetric tensor $F_{\mu\nu}$.
Therefore, it should come as no surprise that ${\cal L}_u$ gives  zero
corrections to $\rho_H$, $p_H$ when applied to the FLRW Universe (where
clearly the antisymmetric tensor vanishes identically $F_{\mu\nu}$) and
hence, higher order counterterms $(u_\mu)^{n}$ in ${\cal L}_u$ are ignored,
while at the same time in ${\cal L}_{cor}$ they have a series of all higher
order derivative terms, up to $n \rightarrow\infty$.
It is easy to check they break energy conservation law and Bianchi
identity, by plugging in the energy conservation equation their expression for $\rho$ and $p$ in
their Eqs.(27) and (28)
\be
\langle \dot \rho \rangle  + 3 H  \langle (\rho + p) \rangle \ne 0
\ee

which we suspect is due to the abovementioned reasons.

Perhaps they could reconstruct their formalism, either by identifying the
correct ${\cal L}_u$  which would be appropriate for the model given by
their ${\cal L}_{cor}$, or by defining an inner product
(following the construction in \cite{corleyjacob}, $( \phi_{in}
\phi_{in})=\int dx \phi F(k)\phi )$
for the fields in such a way that it accounts for the
{\em nonlinear dynamics and Lorentz noninvariance} of short distance modes with
frequency $F(k)$, in an {\em expanding} Universe.
However these issues extend beyond the scope of this paper and we do not
intend to elaborate further.

Acknowledgments: It is a pleasure to thank L. Parker for valuable
discussions during L.M.'s visit in UWM last December. We 
would also like to thank D.Grasso for helpful discussions.

\section{Appendix}

\subsection{Distribution function for the Linear, Crossover and
Transplanckian regime.}

We have a time translation Killing vector for future infinity that
determines our outgoing positive frequency modes.  Let us consider our
Universe as an expanding box, with size $L = a / M $, filled with
modes. Inside this box we have a smaller box with fixed size $l_p
\simeq 1/M$ that determines the range for the transplanckian (TP)
modes. There is a preferred frame attached to the small box (due to
the breaking of Lorentz invariance by the short-distance modes) but
Lorentz invariance is restored in the big box, the expanding Universe.
Following along the arguments and derivation in \cite{corleyjacob} for
the 'particle' distribution function, it is 
straightforward to apply their expression to our model. Below we
consider three regimes depending on the value of the momentum $p$ with
respect to the cuttoff scale $M$

{\bf (a)} $p/M \ll 1$, the ``normal regime''. In this regime the
frequency of the modes becomes linear, \be \omega[p] \simeq p e^{-p/M}
\rightarrow_{p/M\ll1} \;\;\;\;      p\,.  \ee The wavelength of these modes is then
$\lambda \simeq O(L) \gg l_P$.

{\bf (b)} $p/M \simeq O(1)$, the crossover or intermediate regime
during which the TP modes go from the ``TP box'' with fixed size $l_P
=1/M_P$ into the ``big box'' with size $L=a/M$. This process occurs
due to the redshifting of the modes, $p_i=k_i/a$. Each mode $p_i$
will crossover and become a ``normal mode'' at some time $a_i=k_i/M$.

{\bf (c)} $p/M \gg 1$, the TP regime, such that $\lambda \simeq 1/k \ll
l_P=1/M$.

We do not report the derivation of the distribution function since the
reader can find it in detail in \cite{corleyjacob}. In what follows we apply
it to our case, to lend support to our assumption that the TP modes,
{\em modes in the small box with fixed size} $l_p=1/M$ {\em with
respect to the preferred frame}, are not in thermal equilibrium, while
modes in the range of the linear regime, ({\em modes in the big box
with size L=a/M, where Lorentz invariance is restored, the FRW
Universe}) are in a nearly thermal equilibrium situation.

\be {\cal N} (\omega) = \left | \frac{ (\omega_{in}) v_g (p_{in}) }
{(\omega_{out}) v_g (p_{out})} \right | \left |\frac{\beta_p
}{\alpha_p } \right|^2 \,.  \ee The index $in$ ($out$) refers to the
incoming (outcoming) modes as defined in \cite{mbk}; $v_g$ is the
group velocity, \be v_g(p)= \frac{d \omega}{dp} \,, \ee and $\beta_p$,
$\alpha_p$ are the Bogoliubov coefficients, which in our model do not
depend on the momentum $p$ \cite{mbk}. The dispersed frequency is
given in Eq.  (\ref{epstein}), and thus \be |v_g(p_{in})| = \frac
{ \omega_{in}}{p_{in}} \left| 1 - \frac{p_{in}}{M} \right | \,, \ee
and \be v_g(p_{out}) \simeq 1 \,, \ee since $\omega(p_{out}) \simeq
p_{out}$. The phase velocity is defined as $v_c= \omega/p$. Therefore
\be {\cal N} (\omega) = \left | \frac{\beta_p}{\alpha_p} \right|^2 (
e^{-p/M_P}) |e^{-p/M_P} \left ( 1 - \frac{p}{M_P} \right) | \,.  \ee
The thermal distribution is immediately recovered in the limit $p/M
\ll 1$, i.e., $v_g \rightarrow 1$ and $\omega(p) \rightarrow p$.

Now we can evaluate the distribution function for the different
regimes above:

{\bf (a)} In this case \be {\cal N}_a (\omega) \simeq\left
|\frac{\beta_p}{\alpha_p} \right|^2 \,, \ee as it should be since we
recover in the $out$ region the normal linear frequency for the modes,
$p/M \ll 1$ and $v_g \simeq 1$.

{\bf (b)} In the intermediate crossover regime, $p/M \simeq O(1)$, we
have 
\be {\cal N}_b (\omega) \simeq \frac{C}{2}
\frac{e^{-2}}{e^{\beta \omega} -1} \,, 
\ee where
we have identified the inverse of temperature $\beta \simeq a$ and the
linear term in $p/M$ with \be \beta \omega \simeq \frac{1}{1- p/M} = a
\frac{M}{a M - k}\,,\;\;\; \;\;\; \beta \simeq a \,.  \ee Clearly in
this regime $|v_g|$ goes to zero, and $\beta \omega$ goes to
infinity. The spectrum is nearly thermal, however ${\cal N} (\omega)$
goes to zero, since as seen from the TP box the group velocity of
these modes as they approach $p \simeq M$ is becoming zero; or as seen
from the normal particles in the big box, these modes have a high
frequency ($\omega \approx M$) thus they do not contribute very much
to the energy of modes in (a).

{\bf (c)} However in the TP regime the distribution function strongly
deviates from that of thermal equilibrium, since in this case $v_g(in)
\neq 1$, and the frequency is highly nonlinear: 
\be {\cal N}_c
(\omega) \simeq \frac{C}{2} |e^{-p/M}| e^{-p/M} \left | 1 -
\frac{p}{M} \right | \,, \label{nc}
\ee with $p/M \gg 1$. Nevertheless ${\cal
N}_c (\omega) \rightarrow 0$ when $p/M \rightarrow \infty$, thus their
contribution to the energy is suppressed. The suppression comes
directly from the frequency $\omega(p) \simeq p e^{-p/M}$ in this
case.

The volume element in momentum space, $dV_P$, for the dispersed
'particles', whose worldline intersect an hypersurface element
$d\Sigma$ around $x$, having momenta in the range $(p, p+dp)$ is $dV_p
= 2 \delta(p_{\mu}p^{\mu}) dp^4$, where $p$ is future directed. Based
on the definition of \cite{corleyjacob} for the inner product of the
fields $(\phi[_{in}^{out}], \phi[_{in}^{out}] )$ and integrating over
the entire mass shell, the 3-volume in momentum space is given by: \be
dV_3= a^3 \left |\frac{1}{v_g} \right | d^3p \ee which is consistent
with the quantum field theory expressions of currents and 'particle'
number densities, given in terms of creation and annihilation
operators. 

The distribution function, Eq.  (\ref{nc}), justifies our assumptions
of deviation from 
thermal equilibrium in the TP regime.

\subsection{Averaging of the TP energies, pressure and equation of
state.}

We have shown that modes in the TP regime of such very short
wavelength $\lambda_{TP} \ll l_P$ are out of thermal equilibrium. Thus we need
to average their effective contribution to the energy and pressure over many
wavelengths, in order to obtain a nearly thermal, large scale
state. This averaging is done over many wavelengths since 
clearly scales of cosmological order that are of interest to us, are
much much longer than any TP wavelengths. In what follows, we are
interested to find the averaged bare quantities $\langle  \rho_i
\rangle$ and $\langle  \bar{p}_i \rangle$ before including any
modification $\Pi_i$. The viscous pressure modifications,$\Pi_i$,
occur due to the freeze-out and the change in the number of particles
and are accounted for separately in Sect.2. Therefore region II will
be grouped together with region H, since both have a highly dispersed
TP frequency and, for the moment, we are ignoring the freeze-out
corrections. Approximately we can write the energy of regions (0+I) and (II+H) in
one compact form to avoid repetition: \be \rho= p^4 \Theta(M
-p)+\frac{M}{2}p^3 e^{-(p/mp) \Theta(p-M)} \,, \ee where $\Theta(p-M)$
is the unit step function that takes the value $\Theta(p-M)=1$ for $p
> M$ and zero otherwise. Clearly, for modes with $p/M \gg1$ we get
$\rho_{II}$ ($\rho_H$). And for modes $M > p$ we get the radiation
energy density of the (nearly) linear modes.

Using the energy conservation law (while ignoring the freeze-out
effects): \bea \rho + \bar{p}&=& \frac{p}{3} \frac{d \rho}{dp}=
\frac{p}{3} \left(\frac{d \rho_1}{dp}+ \frac{d \rho_2}{dp} \right)
\nonumber \\ & =& ( \rho_1+ \bar{p}_1 ) + ( \rho_2 +\bar{p}_2 ) \,,
\label{p1p2} \eea with \bea \rho_1&=& \frac{M}{2} p^3 e^{-p/M}
\,,\;\;\;\; p > M \,,\\ \rho_2&=& p^4 \,, \;\;\;\; p< M \,.  \eea And
from Eq. (\ref{p1p2}) we find: 
\bea w_1&=& \frac{\bar{p}_1}{\rho_1}=
- \frac{k_B}{3 a M} \Theta(k_B - a M) \,, \\ w_2&=&
\frac{\bar{p}_2}{\rho_2}=\frac{1}{3} \,.  \eea It is clear from the
previous section that since the (nearly) linear modes, $\rho_2$ are
nearly in thermal equilibrium then we do not need to bother with the
averaging procedure for them. (One can however check to verify the
result $\langle w_2\rangle = 1/3$). This is not the case for the modes with $p/M
>1$, since these short wavelengths are out of equilibrium. Let us
calculate $\langle  \bar{p}_1 \rangle$, $\langle  \rho_1 \rangle$, and
$\langle  w_1 \rangle$:

\be \langle  \rho_1 \rangle = \frac{\int_0^a \rho_1 a^2 da}{ \int_0^a
a^2 da} = 3 \frac{\int_0^{a^*=k/M} \rho_1 a^2 da}{ a^3} \simeq
\frac{14 M^4 a^{* 3} }{16 e a^3} \,, \ee

Similarly,

\be \langle  \bar{p}_1 \rangle = \frac{\int_0^a \bar{p} a^2 da}{
\int_0^a a^2 da} = 3 \frac{\int_0^{a^*=k/M} \bar{p} a^2 da}{ a^3}=
-\frac{14 a* }{20 a} \langle  \rho_1 \rangle \,.  \ee

and \be \langle  w_1 \rangle = \frac{\langle  \bar{p}_1 \rangle}{\langle 
\rho_1 \rangle } = 0 - \frac{14 a^*}{20 a} \simeq - \frac{B}{a} \,,
\ee where in Planck units, $a^* = l_p$ and we have used the
normalization that Planck length $l_p = 1/M= a^*= a(t_P)=1$. The time
$a^*$ corresponds to the crossover time when the wavelength of the TP
modes is on average of order the size of the small box, $p(a^*)=M$.
This time scale of order the Planck box is much smaller than scales of
cosmological interest, $L=a/M$. External observers bound to the large
scale, Lorentz invariant Universe with size $L=a/M$ and in thermal equilibrium, 'feel' the
energy and pressure contributions from the TP modes given by Eq. (\ref{nc}).

Of course the real equation of state for these modes is given by their
effective equation of state, $\tilde w_i$, Sect.2, that takes into
account the relativistic kinetic theory modifications due to the
freeze-out.

In a similar manner, one can obtain $\langle w_2\rangle$ and in particular the
numerical coefficient $A$ for $\langle w_I\rangle, \bar p_I, \rho_I$.

\end{document}